# NUMERICAL ANALYSIS OF THE QUARK FRACTION OF THE PROTON SPIN*

JEFFREY E. MANDULA

*Department of Energy, Division of High Energy Physics, Washington, DC 20585*

We report on a lattice QCD estimate of the quark spin fraction of the proton spin. The estimate is arrived at by means of a lattice QCD simulation of the polarized proton matrix element of the Adler-Bell-Jackiw anomaly. The preliminary result of the simulation is that this fraction is rather small. This is in accord with the interpretation of the EMC experiment that the quark spins are responsible for very little, if any, of the proton spin.

## 1. Introduction

In this talk I will report on a recent lattice QCD calculation of the fraction of the proton's spin that is carried by the spins of its constutent quarks. This analysis was inspired by the famous measurement of the spin structure of deep inelastic muon scattering from protons carried out by the European Muon Collaboration (EMC)[1]. Their principal result, which has stirred a great deal of interest, is

$$\int_0^1 g_1(x)\, dx = .126 \pm .010 \pm .015 \quad (1)$$

The quoted systematic error of $\pm .015$ includes an estimate of the uncertainty due to the need to extrapolate the measured value of the structure functions to the regions above $x = .7$ and below $x = .01$.

As has been said many times before, the significance of the EMC measurement comes from combining it with information from neutron β-decay and strangeness changing hyperon decay. With the aid of isospin and flavor SU(3), one obtains for the individual quark spin fractions

$$\begin{aligned} \Delta u &= +.74 \pm .05 \\ \Delta d &= -.51 \pm .05 \\ \Delta s &= -.19 \pm .07 \end{aligned} \quad (2)$$

and for the total quark spin fraction

---

$$\Delta\Sigma = \Delta u + \Delta d + \Delta s = .04 \pm .16 \tag{3}$$

The conclusion that only a small fraction, if any, of the proton's spin is carried by the spins of its constituent quarks was completely unanticipated, as was the conclusion that the strange quark's contribution was comparable to that of the down quark. This result is a challenge to QCD, simply because it is so much at variance with the expectations based on the parton model, which usually provides reliable guidance. On the other hand, it is also an opportunity for QCD to show its mettle by accounting quantitatively for such a surprising result.

## 2.  A "Gluonic" Contribution

The EMC result led to a very interesting attempt to attribute the apparently missing spin to a cancellation between intrinsic quark contributions and an indirect gluonic contribution. The idea is that the same diagram that is responsible for the Adler-Bell-Jackiw anomaly[2] (Figure 1), also contributes to the quark spin fractions, which are matrix elements of the quark axial current operators between polarized proton states. This idea has been criticised because it has proved difficult to phrase it in a gauge invariant manner. This is related to the fact that, from an operator product expansion perspective, the only operators which can contribute to the zeroth moment of $g_1$, Eq. (1), are the quark axial current operators[3]. There are no gauge invariant gluon field operators with the correct quantum numbers. There is, however, a gauge variant operator, the anomalous current $K_\mu$, whose divergence is the anomaly. The analysis of Efremov and Turyaev, Alterelli and Ross, and Carlitz, Collins, and Mueller[4] indicates that if this contribution is quite substantial, it could be responsible for the apparently fairly large (negative) value of the strange quark spin fraction, and would make a large negative contribution to the total spin fraction. The smallness of the total would then be due to a cancellation between this and a respectably large positive "intrinsic" quark contribution.

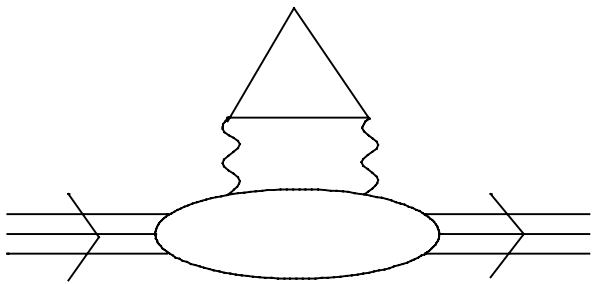

**Figure 1**  The triangle diagram, which is responsible for the Adler-Bell-Jackiw anomaly.

In order to see if this idea is on the right track, a numerical evaluation of this hidden gluonic contribution was carried out on the lattice[5]. It was, of course, neccessary to work in a specific gauge, and that chosen was the temporal axial gauge. The lattices were a collection of rather small ones which had been generated by Bernard, Soni, and their collaborators[6] in some early work on the computation of weak matrix elements.

We will not review here the analysis of the numerical uncertainties associated with the computation (which could easily be a factor of 2), but only summarize the result. Within the statistical errors of the computation, the gluonic contribution was too small to observe. If the statistical errors are taken as a bound on the value, then the modification of each of the Eqs. (2) is less than .02, and the modification of $\Delta\Sigma$, Eq. (3), is less than .05. This is an order of magnitude too small to affect the thrust of the EMC result.

### 3.  The Total Quark Spin Contribution

The Adler-Bell-Jackiw triangle anomaly also provides an indirect method for evaluating $\Delta\Sigma$ using lattice QCD. It gives a computationally feasible calculation which can be directly compared with the experimental results. For each flavor of quark, in the chiral limit,

$$\partial_\mu j_\mu^5 = \partial_\mu \bar{\psi} i\gamma_\mu \gamma_5 \psi = \frac{\alpha}{2\pi} \varepsilon_{\mu\nu\lambda\sigma} Tr F_{\mu\nu} F_{\lambda\sigma} \qquad (4)$$

The idea is to calculate the forward matrix elements of the spatial components of $j_\mu^5$ by computing matrix elements of $\partial_\mu j_\mu^5$ between states with different momenta, dividing by the momentum transfer, and then extrapolating to zero momentum transfer. The anomaly condition is used to express $\partial_\mu j_\mu^5$ in terms of link variables, since, in a simulation, the expression of $F_{\mu\nu}\tilde{F}_{\mu\nu}$ does not involve the solution of the Dirac equation lattice by lattice. In this way, the calculation becomes comparable in computational difficulty to the calculation of a glueball propagator, rather than to the computation of weak matrix elements.

### 4.  Lattice Analysis

The implementation of this idea is straightforward. The total, flavor singlet, quark spin fraction is explicitly given by

$$2m s_i \Delta\Sigma = \langle \vec{p}=0\ s | \sum_{flavors} j_i^5 | \vec{p}=0\ s \rangle \qquad (5)$$

where $s_i$ is the polarization vector of the proton state. From the divergence of $j_\mu^5$ we can express this as

$$2ms_i\Delta\Sigma = \lim_{p\to 0} \frac{1}{2ip} \langle \vec{p}s | \sum_{flavors} \partial_\mu j_\mu^5 | -\vec{p}s'\rangle$$

$$\vec{p} = p\hat{i} \parallel \vec{s} \parallel \vec{s}'$$

$$(E,\vec{p})\cdot s = (E,-\vec{p})\cdot s' = 0$$

(6)

To evaluate this on the lattice we make the approximation of working in the chiral limit, and use

$$\sum_{flavors} \partial_\mu j_\mu^5 = \frac{N_{flavors}\alpha}{2\pi}\varepsilon_{\mu\nu\lambda\sigma} Tr\, F_{\mu\nu} F_{\lambda\sigma}$$

(7)

To express this in terms of link variables we use the symmetrical lattice definition of the color field strength[7]

$$F_{\mu\nu}^{latt}(n) = \frac{1}{4}\left[ U_\mu(n) U_\nu(n+\hat{\mu}) U_\mu^\dagger(n+\hat{\nu}) U_\nu^\dagger(n) + 3\ symmetrical\ terms \right]$$

$$-\ Hermitean\ Conjugate\ -\ Trace$$

(8)

where the symmetry is as indicated in Figure 2. Since under lattice gauge transformations, the change in this expression only depends on the value of the gauge transformation at the central site, the resulting expression for the anomaly is gauge invariant, as it should be.

On a finite lattice we cannot actually reach the zero momentum transfer limit, since there is a finite minimum momentum. We therefore use the expression for $\Delta\Sigma$ at the lowest two values of the momentum. As in all lattice calculations of physical matrix elements, we must extract the matrix element by taking the large time limit of Euclidean Green's functions. One must also take care to remove the actual, momentum dependent, falloff from the Green's function. This is accomplished by using an expression for $\Delta\Sigma$ which would give the exact value in the infinite time continuum limit, and which further has the property that the finite spacing and finite time errors cancel as completely as possible. Specifically, we use as a lattice approximation to $\Delta\Sigma$ the expression

$$\Delta\Sigma = \frac{N_f \alpha}{2\pi} \frac{Tr\, P\, \langle 0|\Psi_{\vec{p}}(x_4) F\tilde{F}(y_4) \bar{\Psi}_{-\vec{p}}(z_4)|0\rangle}{i|\vec{p}|\, Tr\, P_+ \langle 0|\Psi_{\vec{p}}(x_4) \bar{\Psi}_{\vec{p}}(z_4)|0\rangle} \qquad (9)$$

$$P = P_+ P_s P_{s'}$$

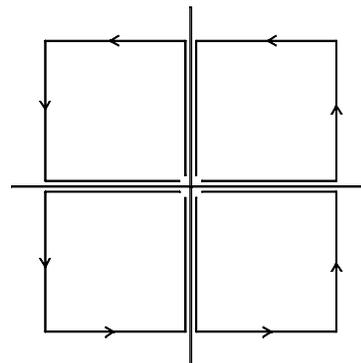

**Figure 2** Symmetrical structure of $F_{\mu\nu}^{latt}$

The color singlet proton field $\Psi$ is formed from colored quark fields by projecting on the appropriate spin and symmetry. The matrices $P_+$ and $P_s$ project on positive energies and on the indicated polarization. This expression is convenient in that the normalization of the proton field cancels out.

It is crucial in any calculation of physical quantities to understand as well as possible the inherent errors, those which are non-statistical. Three sources of such errors in the present calculation are the finite lattice spacing, the fact that chiral symmetry is not exact (*i.e.* the quark bare masses are not zero), and the need to extrapolate a minimum finite momentum. All these are reduced, but none eliminated, by averaging over opposite values of the finite spatial momentum. From dimensional considerations, the order of magnitude of the errors coming from these sources is as follows:

| Physical Origin | Order of Magnitude | $(\vec{p}) + (-\vec{p})$ Symmetrized |
|---|---|---|
| Chirality Breaking | $\dfrac{m_{quark} a \Lambda_{QCD}}{p_{latt}}$ | $m_{quark}/\Lambda_{QCD}$ |
| Finite Spacing | $a\Lambda_{QCD}^2 / p_{latt}$ | $a\Lambda_{QCD}$ |
| $p_{latt} \neq 0$ Extrapolation | $p_{latt}/\Lambda_{QCD}$ | $(p_{latt}/\Lambda_{QCD})^2$ $\left(p_{latt} = \dfrac{2\pi}{aN_{side}}\right)$ |

An additional source of error is fact that the proton propagator receives contributions from higher mass states with the same quantum numbers as the proton. At infinite time, of course, the single proton controls the propagator. This contribution is a simple exponential (actually a hyperbolic cosine with periodic boundary conditions), and so the actual propagator on the lattice allows one to see to what extent the single proton contribution dominates at any given value of the Euclidean time separation. The lattice proton propagator at the three smallest values of the spatial momentum on the $16^3 \times 24$ lattices, $k = 0, 1, 2$ times $2\pi/aN_{side}$, is shown in Figure 3. The single proton contribution is identified by extrapolating from large times. As one can see, high mass states are important below 5 units of Euclidean time. The state consisting of a single proton is not dominant until $t > 5$. The effect of the higher mass states should be to decrease the apparent value of $\Delta\Sigma$. The reason is that although these states decay away relative to the single proton contribution eventually, at finite time they will have the effect of diluting the proton contribution.

## 5. The Loop Anomaly on Quenched Lattices

The effect of quenching on the anomaly is rather subtle and complicated. The quenched approximation includes one-quark-loop diagrams (Figure 1), corresponding to the contraction of the two quark operators in axial current. Thus the anomaly is present in the quenched approximation. However, its value may be different than in the full theory. The reason for this lies in the $\eta'$.

In the full theory, the $\eta'$ has a finite mass in the chiral limit, and so it does not make any singular pole contribution at vanishing momentum transfer. However, in the quenched approximation, the $\eta'$ mass is zero, and this gives an extra contribution to the matrix elements of the anomaly. Figure 4 shows the connection between quenching and the $\eta'$ mass. It is exactly the iterated bubble diagrams, absent in the quenched approximation, that lift the degeneracy between the $\eta'$ and the $\eta$, which is a Goldstone boson in the chiral limit.

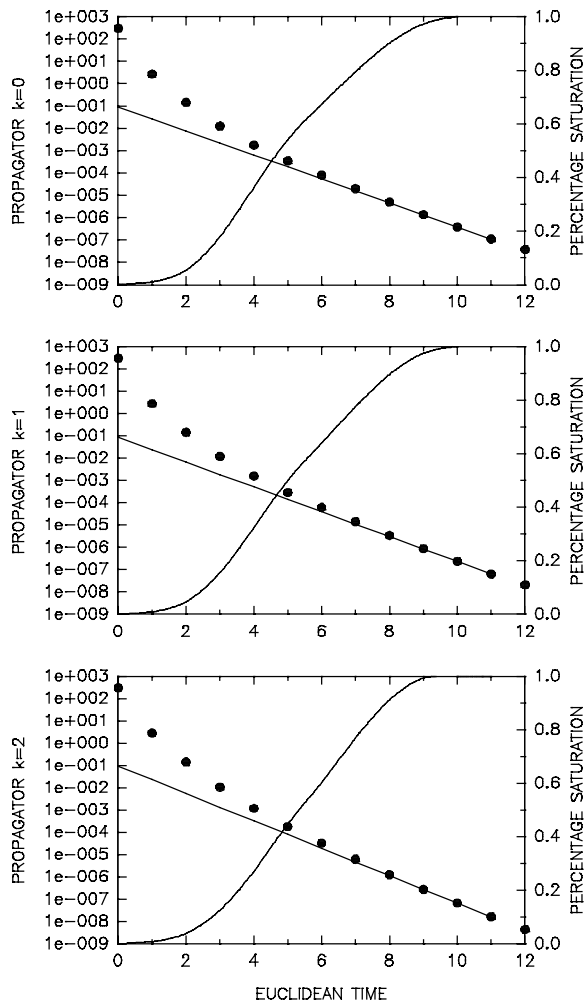

**Figure 3** The saturation of the proton propagator by the contribution of the single proton state

The single-quark-loop diagrams are singular at zero momentum transfer, and these singularities contribute to the matrix elements that concern us, as well as to other related quantities. Specifically, there are "extra" contributions to matrix elements of the flavor singlet axial current, the pseudoscalar density, and the anomaly due to the zero mass of the $\eta'$ in the chiral limit. These are double and single poles in $\langle p|j_\mu^5|p\rangle$ double and single poles in $\langle p|j^5|p\rangle$ and a single pole in $\langle p|F\tilde{F}|p\rangle$ The double poles cancel, but the remaining contributions depend on order of chiral and $q \to 0$ limits. In forward matrix elements of the anomaly and pseudoscalar density, finite residual contributions due to the $\eta'$ are present in the zero quark mass limit. On the other hand, in the zero quark mass case, the $\eta'$ produces $q \to 0$ divergent terms in both the anomaly and the axial current. The upshot of these remarks is that dynamical quarks probably needed, and certainly preferable, to assure smooth chiral and $q \to 0$ limits.

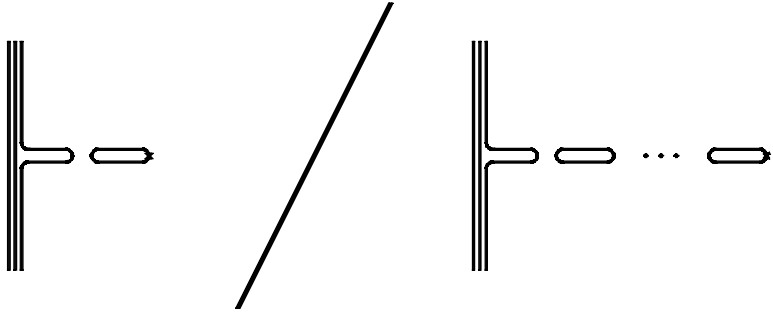

**Figure 4** Diagrammatic description of multiple closed quark loops giving a non-zero $\eta'$ mass

## 6. Results

In the remainder of this talk, we report on progress towards arriving at a quantitative estimate of $\Delta\Sigma$. The calculation used lattices generated by Bernard, Hockney, and Soni for the calculation of hadronic matrix elements that enter into weak interaction rates. They made available for the present calculation two ensembles of lattices, both at $\beta = 5.7$. Along with the lattices were Wilson quark propagators. One was a set of 204 $6^3 \times 10$ lattices with propagators computed through 8 units of Euclidean time both forward and backwards. This is the same set that was used in the gauge variant calculation briefly reviewed in Section 2. The other was a set of 16 $16^3 \times 24$ lattices with similar propagators computed through 12 units of Euclidean time. The quark propagators satisfy open (Neumann) boundary conditions with hopping constants $\kappa = .162$ for the small lattices and $\kappa = .160$ for the larger ones.

To keep the systematic errors as small as possible, we work at the largest Euclidean time separations, avoiding only the endpoint of the propagator because it is likely to be distorted by the boundary condition. On the small lattices we only use the lowest value of the lattice spatial momentum. We keep three values of the time separation and forward and backward propagation separate to confirm that the data are consistent. Table I shows results of the simulation of $\Delta\Sigma$.

**Table I** Simulation of $\Delta\Sigma$ on $6^3\times10$ lattices

| Separations | | $\Delta\Sigma$ | |
|---|---|---|---|
| $x_4 - y_4$ | $y_4 - z_4$ | Forward | Backward |
| 4 | 3 | $-.033 \pm .103$ | $+.003 \pm .032$ |
| 3 | 4 | $+.116 \pm .143$ | $+.003 \pm .029$ |
| 3 | 3 | $-.006 \pm .057$ | $-.003 \pm .018$ |

On the larger lattices, we average all values of the time separations with minimum propagation time 5, to boost the statistical precision. We keep separate as a consistency check the three directions of the spatial momentum as well as keeping separate the forward and backward propagation. The results, along with the average of the three separate forward and backward propagation results are shown in Table II. The errors shown in the tables are the statistical errors only.

The intrinsic errors are substantial on these lattices. On both sets of lattices the quantity which characterizes chirality, $m_{quark}/\Lambda_{QCD}$ is about 1. The momentum extrapolation is characterized by $p_{latt,min}/\Lambda_{QCD}$. On the small lattices this is also about 1, while on the larger ones it is somewhat under 1/2.

### 7. Conclusions

The results presented here are clearly very preliminary. The statistical errors are larger than the signal, which means that we have a bound on the magnitude of $\Delta\Sigma$. Based on the statistical errors that bound is about 1/4. Two of the systematic errors which seem quite large are the quark mass and the dilution by higher mass states with proton quantum numbers. The former calls for larger values of the hopping constant, and probably an extrapolation to zero mass if the statistical errors become small enough. The latter calls for longer Euclidean time propagation. Finally, the calculation should be attempted on lattices generated using full QCD, not the quenched approximation we have reported on here.

**Table II** Simulation of $\Delta\Sigma$ on $16^3\times24$ lattices

| Direction | $\Delta\Sigma$ | |
|---|---|---|
| | Forward | Backward |
| 1 | $+.257 \pm .140$ | $+.253 \pm .277$ |
| 2 | $-.057 \pm .213$ | $+.103 \pm .140$ |
| 3 | $-.373 \pm .098$ | $-.068 \pm .199$ |
| Average | $-.058 \pm .109$ | $+.096 \pm .123$ |

- 9 -

Although we have couched our results in many caveats, which are certainly appropriate, the results obtained so far indicate that $\Delta\Sigma$ is quite small. QCD seems to be on the right track to account for the surprising EMC result.

**Acknowledgements**